\documentclass{article}

\usepackage[utf8]{inputenc}
\usepackage{tismir}
\usepackage{amsmath}
\usepackage{hyperref}
\usepackage{url}
\usepackage{graphicx}
\usepackage{booktabs}
\usepackage{lipsum}
\usepackage{makecell}
\usepackage{multirow}
\usepackage{endnotes}
\usepackage{svg}
\usepackage{float}

\usepackage[utf8]{inputenc}
\usepackage{csquotes}
\usepackage[english]{babel}
\usepackage[
    style=apa,
    sortcites=true,
    backend=biber,natbib,
    labeldate=year
]{biblatex}
\AtEveryBibitem{\clearfield{month}}
\AtEveryBibitem{\clearfield{day}}
\DeclareNameAlias{sortname}{family-given}
\DeclareNameAlias{default}{family-given}

\usepackage{tikz}
\def\checkmark{\tikz\fill[scale=0.4](0,.35) -- (.25,0) -- (1,.7) -- (.25,.15) -- cycle;} 

\usepackage{caption}
\usepackage{varwidth}
\DeclareCaptionFormat{capfmt}{%
  \begin{varwidth}{\linewidth}%
    \centering
    #1#2#3%
  \end{varwidth}%
}

\title{Automatic Identification of Samples in Hip-Hop Music via Multi-Loss Training and an Artificial Dataset}

\newcommand*\samethanks[1][\value{footnote}]{\footnotemark[#1]}
\author{
    Huw Cheston\thanks{Centre for Music \& Science, University of Cambridge. Correspondence should be addressed to: \href{mailto:hwc31@cam.ac.uk}{Huw Cheston, hwc31@cam.ac.uk}} \protect\endnote{The first author completed this work during a research internship at Spotify.}, 
    Jan Van Balen\thanks{Spotify}, 
    Simon Durand\samethanks
}

\date{}

\type{research}

\authorref{Cheston,~H., Van Balen,~J., and Durand,~S.}

\authorshort{Cheston,~H., Van Balen,~J., and Durand,~S.}
\titleshort{Automatic Identification of Samples in Hip-Hop}

\begin{document}


\twocolumn[{%
\maketitleblock
\begin{abstract}
Sampling, the practice of reusing recorded music or sounds from another source in a new work, is common in popular music genres like hip-hop and rap. Numerous services have emerged that allow users to identify connections between samples and the songs that incorporate them, with the goal of enhancing music discovery. Designing a system that can perform the same task automatically is challenging, as samples are commonly altered with audio effects like pitch- and time-stretching and may only be seconds long. Progress on this task has been minimal and is further blocked by the limited availability of training data. Here, we show that a convolutional neural network trained on an artificial dataset can identify real-world samples in commercial hip-hop music. We extract vocal, harmonic, and percussive elements from several databases of non-commercial music recordings using audio source separation, and train the model to fingerprint a subset of these elements in transformed versions of the original audio. We optimize the model using a joint classification and metric learning loss and show that it achieves 13\% greater precision on real-world instances of sampling than a fingerprinting system using acoustic landmarks, and that it can recognize samples that have been both pitch shifted and time stretched. We also show that, for half of the commercial music recordings we tested, our model is capable of locating the position of a sample to within five seconds. \\

\textbf{Words in abstract: } 231
\end{abstract}
\begin{keywords}
Sample Identification, Audio Retrieval, Artificial Dataset, Hip-Hop, Convolutional Neural Networks\\
\end{keywords}
}]
\saythanks{}

\section{Introduction}\label{sec:headings}

In music production, sampling refers to the reuse of recorded music, sounds, or dialogue from another source to create a new work. Sampling originated as a creative practice in both the tape collages of musique concrète and the riddims of Jamaican dancehall, but became predominantly associated with hip-hop music in the 1970s \citep{chang_cant_2011}. Nowadays, it is widespread in many mainstream music genres, with 17\% of songs charting on the 2022 Billboard Hot 100 sampling from previously released songs.\endnote{\href{https://www.tracklib.com/blog/state-of-sampling-2022}{https://www.tracklib.com/blog/state-of-sampling-2022}} However, the association between sampling and hip-hop music still remains strong \citep{schloss_making_2013}.

Information relating to the origin and reuse of hip-hop samples can enhance music understanding, discovery, and recommendation. Users of numerous music streaming platforms have created playlists of songs that all feature the same sample, such as the drum break from the song ``Amen, Brother''. Both the \href{https://www.whosampled.com/}{WhoSampled} and \href{https://secondhandsongs.com/}{SecondHandSongs} services allow users to browse individual songs by the samples that they incorporate, and discover other songs featuring the same samples. Neither of these services are automated; rather, data is contributed by a community of users (who manually identify the relationships between two recordings), and subsequently reviewed by moderators.

A limited number of attempts have been made to perform this task of \textit{sample identification} automatically, with the goal being to design a system capable of detecting whether a query song samples from a pool of potential candidate recordings. Parallels thereby exist with tasks such as audio fingerprinting and cover song identification. Using the definitions provided by \citet{lee_similarity-based_2021}, fingerprinting is a high specificity retrieval task, requiring the same music track as the query to be retrieved from a database. Meanwhile, cover song identification is of a lower specificity, where other versions of the same underlying musical work as the query are retrieved. 

Like audio fingerprinting, in sample identification the candidate audio is contained within the query recording. However, like cover song identification, a candidate sample is typically transformed in some way within a query recording. Producers may apply audio effects to a sample in order to change its pitch, timbre, volume, or structure, and may also layer it against new recordings or other samples. Further adding to the difficulty of the sample identification task is that only a small part of a candidate may be relevant to a query, such as when a few seconds of audio are sampled and looped repeatedly \citep{schloss_making_2013}. This effectively limits both the rate and total amount of discriminative information between a candidate sample and query recording. Perhaps as a result, so far minimal progress has been made on this task.

A small number of previous approaches to sample identification in hip-hop music have drawn from audio signal processing algorithms. \citet{vanbalen_sample_2013} optimized an existing fingerprinting system based on extracting prominent spectral ``landmarks'', consisting of hashed tuples of time-frequency peaks \citep{wang_industrial-strength_2004}. This algorithm is not able to recognize audio that has been either pitch-shifted or time-stretched; \citet{vanbalen_sample_2013} accounted for this by re-scaling the landmarks extracted from a query music track and computing the best match with all candidate samples. Overcoming this limitation, \citet{sonnleitner_landmark-based_2016} proposed the use of landmarks with four peaks --- which are robust to both rotation and isotropic scaling --- to fingerprint the music recordings played in DJ mixes, which may also be pitch-shifted or time-stretched. The use of audio landmarks combined with text and video features has also been proposed by \citet{smith_classifying_2017} to detect derivative musical works on video sharing websites, with instances of sampling mentioned as another possible application by the authors. 

Two attempts at this task have used non-negative matrix factorization (NMF), with this first outlined as a possible strategy for sample identification by \citet{dittmar_audio_2012}. \citet{whitney_automatic_2013} used NMF to obtain pairwise cross-correlations between a candidate sample and windowed sections of a query music track. They accounted for both pitch shifting and time stretching by performing multiple NMFs with resampled versions of the audio obtained by taking the ratio of the sample and song tempo. \citet{gururani_automatic_2017} also used dynamic time warping (DTW) to compute the alignment between NMF activations of a query song and candidate sample, extracting handcrafted feature vectors from this path and using them as the input to a binary classifier. They accounted for pitch shifting and time stretching by computing multiple NMFs with the audio transposed by differing amounts, with the most likely pitch candidate being the one that minimized the cost of the DTW function required to map the query onto the candidate. The success of NMF in identifying samples that have been heavily transformed by audio effects remains unclear, however.

In contrast with both acoustic landmarks and NMF, modern approaches to fingerprinting and cover song identification typically leverage advances in deep learning to train systems end-to-end. Convolutional neural networks (CNNs) have proven capable of generating embedded representations of musical audio that are invariant to changes in spectral content and musical attributes \citep{chang_neural_2021, du_bytecover_2021, xu_key-invariant_2018}. Consequently, they are appropriate both in fingerprinting, where background noise and room reverberation might be present, and also cover song identification, where the musical key, tempo, and genre may change between versions. CNN-based approaches have outperformed conventional, non-neural approaches to both tasks, improving on spectral landmarks in fingerprinting \citep{chang_neural_2021} and DTW-based sequence alignment methods in cover song identification \citep{xu_key-invariant_2018}. However, and despite the similarities between these tasks and sample identification, to the best of our knowledge no published research has considered training an end-to-end, deep sample identification model.

In this paper, we adapt a CNN architecture previously used for cover song identification by \citet{du_bytecover_2021} to the task of identifying samples in hip-hop music. We bypass the limited availability of commercial training data by training our model on an artificial dataset: ``samples'' are created by extracting stems from non-commercial music recordings using a source separation algorithm, and the model is then trained to classify whether a music recording contains a subset of these stems, transformed using various digital audio effects. This has several advantages over training on real-world instances of sampling, insofar as it: (1) minimizes the probability of overfitting to particularly common samples; (2) enables nearly any music recording to be used in training, regardless of if it contains a sample; (3) allows the degree of transformation applied to a sample to be explicitly controlled.

\begin{figure*}[t]
  \centering
  \includegraphics[width=0.85\textwidth]{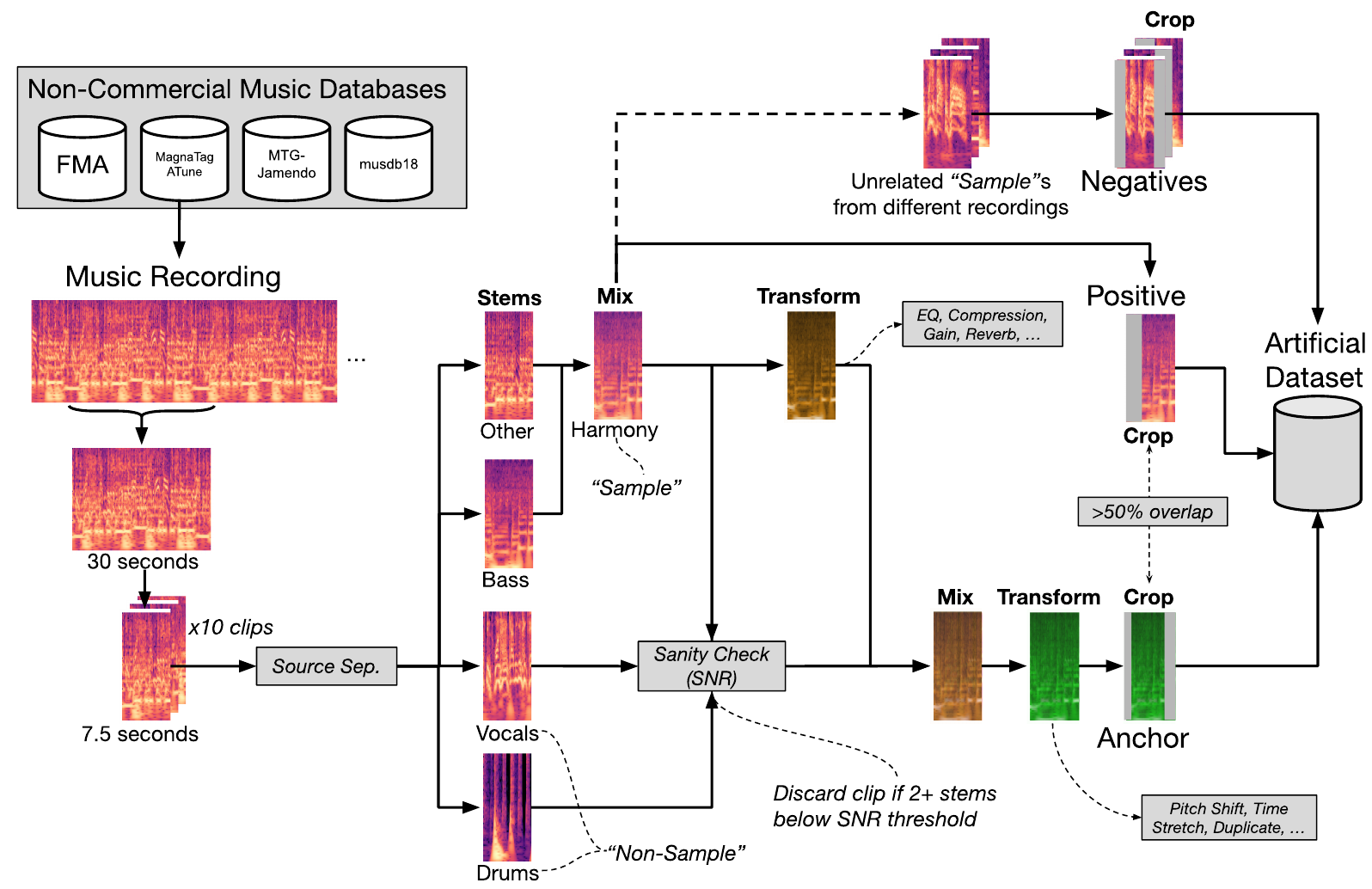}
  \caption{The artificial dataset used during training. For the music recording shown here, $N = 3$ stems passed the sanity (SNR) check, with the \textit{harmony} stem used as the ``sample'' and the \textit{vocals} and \textit{drums} as ``non-sample''. Note that the selection of stems that make up the sample and non-sample are randomized in practice.}
\label{fig:artificial_dataset}
\end{figure*}

We begin by outlining the construction of both our artificial dataset and our proposed system, the performance of which we evaluate on a dataset of commercial hip-hop music recordings. We describe the results of several experiments and compare our model with several baseline audio fingerprinting systems, showing that our approach outperforms them in terms of retrieval precision. Finally, we demonstrate that our model generalizes to locating the section of audio that is sampled in a given commercial recording.

\section{Methods}\label{sec:methods}

\subsection{Artificial Dataset}\label{sec:artificial_dataset}

A diagram showing how we constructed our artificial dataset is given in Figure \ref{fig:artificial_dataset}.

\subsubsection{Audio Sources}\label{sec:audio_sources}

To create our artificial dataset, we use non-commercial music recordings from four existing open access databases: (1) the Free Music Archive (FMA) \citep{defferrard_fma_2017}, (2) MagnaTagATune \citep{law_evaluation_2009}, (3) MTG-Jamendo \citep{bogdanov_mtg-jamendo_2019}), and (4) musdb18 \citep{rafii_musdb18_2017}. We collect all recordings tagged with ``hip-hop'' or ``rap'' either by the artists themselves (FMA, musdb18) or by human annotators (MagnaTagATune, MTG-Jamendo).\endnote{For FMA, we include tracks where \texttt{genre\_top == ``Hip-Hop''}. For MagnaTagATune, we include tracks annotated by human players of the ``TagATune'' as either \texttt{rap} or \texttt{hiphop}. For MTG-Jamendo, we include tracks where \texttt{genre\_dortmund == ``raphiphop''}, \texttt{genre\_rosamerica == ``hip hop''}, or \texttt{genre\_tzanettakis == ``hip hop''} for at least one of three human annotators. For musdb18, we include tracks where \texttt{Genre == ``Rap''} in the provided \href{https://github.com/sigsep/website/blob/master/content/datasets/assets/tracklist.csv}{tracklist}.} This results in 3,804 tracks, to which we add a random subset of 31,196 additional recordings from FMA tagged with ``Rock'', ``Electronic'', ``Pop'', ``Jazz'', or ``Soul-RnB''. Recordings from these genres are all commonly used as samples in rap and hip-hop \citep{vanbalen_sample_2013, chang_cant_2011}.

Across these 35,000 music recordings, we discard 530 (1.5\%) which had a duration of less than thirty seconds. The median duration of those remaining is \texttt{03:27} minutes:seconds, with the longest track lasting for over five hours. Consequently, we trim a thirty-second excerpt from every recording, beginning at either \texttt{00:30} (for the 76.1\% of total recordings that were longer than one minute) or \texttt{00:00} (for all remaining recordings). We downmix these audio excerpts to mono with a sample rate of 22,050 Hz.

We choose thirty seconds as the duration for our audio excerpts as it is approximately equivalent to the length of the longest sample in our dataset of commercial audio (see section \ref{sec:evaluation_dataset}). While it would be possible to use more than one excerpt per song --- or to increase the duration of the single excerpt --- we argue that, given the loop-based, electronic and hip-hop music that appears most frequently in the artificial dataset, this would be unlikely to significantly increase the variety of the training audio, over and above simply adding more source music tracks to the dataset.

\subsubsection{Dataset Curation}

Given the types of user-submitted material contained in our source databases, using every one of these excerpt would likely cause non-music recordings to be included in the artificial dataset. Subsequently, we aim to remove recordings that contained activity in only a restricted number of musical elements; for instance, tracks that only contained vocals could feasibly be spoken word or interviews, while percussion-only recordings could be drum breaks or sound effects.

We use the open-source \texttt{demucs} source separation algorithm \citep{rouard_hybrid_2022} to extract four stems from each audio excerpt: \textit{vocals}, \textit{bass}, \textit{other}, and \textit{drums}. Following \citet{huang_modeling_2021}, we mix the \textit{bass} and \textit{other} stems together by summing them, as both typically fulfill a similar role as harmonic accompaniment. We refer to this combined stem as the \textit{harmony} stem. 

In order to estimate the number of stems that did not contain activity, we calculate the signal-to-noise ratio (SNR) for each of the three extracted stems, using the sum of the remaining two stems as the ``signal''. We remove excerpts from the artificial dataset where at least two of the three stems fall below a threshold which, after listening to the properties of the extracted stems, we set to $-20$ dB SNR.

We include the 29,210 excerpts that pass this preprocessing stage (83.5\% of total recordings) in the artificial dataset. For 16,482 of these excerpts (56.4\%), $N = 3$ stems are above the SNR threshold, with $N = 2$ for all remaining tracks. The proportion of individual stems in these recordings that are above the SNR threshold was: \textit{vocals}, 77.0\%; \textit{harmony}, 97.9\%; \textit{drums}, 81.4\%. We split these excerpts into artificial training and validation sets in the ratio 4:1.

As part of our supplementary materials, we include several examples of recordings that failed the SNR check --- the majority of which are either non-music recordings or are music recordings consisting of only a reduced number of instruments (e.g., solo guitar, unaccompanied piano).\endnote{This will be released at a later date.}

\subsubsection{Window Extraction}\label{sec:window_extraction}

Previous cover song identification models have used large sections of musical audio as input, up to and including an entire recording \citep{yesiler_accurate_2020, du_bytecover_2021, xu_key-invariant_2018}. This is unlikely to be helpful in the sample identification task, where only a small portion of a candidate may be relevant to a query \citep{schloss_making_2013}. As such, we use short windows of audio during training, similar to the ``shingles'' used in remix detection by \citet{casey_fast_2007}. 

For a window with $N$ stems above the SNR threshold (where $N\in\{2,3\}$), we treat any combination of stems up to and including $N - 1$ as the ``sample''. All other stems are considered ``non-sample''. For tracks where all three stems exceed the SNR threshold, the ``sample'' could thus be any one of the \textit{vocals}, \textit{harmony}, or \textit{drums} stems, or the combination of two of these stems (i.e., \textit{harmony} $+$ \textit{drums}). For tracks where two stems exceed the threshold, only one of these is treated as the ``sample'', with the other as the ``non-sample''. Note that this approach assumed that samples always appear with additional elements (represented by the ``non-sample'' stems) layered against them. We test the alternative possibility --- that samples might be used wholesale by a producer with no additional layering --- in our experiments.

We set the initial window size to 7.5 seconds with a hop of 2.5 seconds between windows, meaning that a single excerpt from the artificial database yields ten windows. Later in the data transformation process, we truncate every window to five seconds (see section \ref{sec:audio_fx}). This value was chosen following the work of \citet{gururani_automatic_2017}, which indicated an average duration of 4.5 seconds in their database of hip-hop music samples. 

For each window, we include every possible combination of ``sample'' and ``non-sample'' stems in the artificial dataset. Consequently, a single thirty-second audio excerpt with $N = 3$ stems above the SNR threshold yields 60 windows. In total, 989,184 windows are included in the artificial training set (1,375 hours of audio), and 241,664 in the validation set (336 hours). Note that all of the windows derived from a single music recording appear in the same split of the dataset.

\subsubsection{Audio Transformation}\label{sec:audio_fx}

We use pairs of audio clips during training, each of which comprises an anchor and a positive clip. The positive clips are simply the original ``sample'' stems, which we sum together to create a single mono audio file. The ``anchor'' clips consist of both the ``sample'' and ``non-sample'' stems, to which we apply audio transformations using the \texttt{pedalboard} Python library \citep{sobot_pedalboard_2021}. In a real-world sample identification set-up, these represent the candidate (or ``sample'') and query (or ``song'') tracks, respectively.

\begin{table*}[t!]
    \centering
        \begin{tabular}{c|c|c|c|c}
        \toprule
            \textbf{FX name}&  \textbf{Parameters}&  \textbf{FX} \(\in\mathbb{A}\)&  \textbf{FX} \(\in\mathbb{B}\)& \textbf{FX} \(\in\mathbb{C}\)\\ \hline
            Band EQ&  \makecell{$N$(bands) \(\sim\boldsymbol{U}(1, 8)\), \textbf{then}: \\ \textit{Band} gain \(\sim\boldsymbol{U}(-20, 10)\) dB \\ \textit{Band} frequency \(\sim\boldsymbol{U}(50, 11025)\) Hz \\ \textit{Band} Q \(\sim\boldsymbol{U}(0, 1)\)} &  & \checkmark & \checkmark \\ \hline
            Compression&  \makecell{Threshold \(\sim\boldsymbol{U}(-30, 0)\) dB \\ Ratio \(\in\{2, 4, 8, 20\}\) dB \\ Attack \(\sim\boldsymbol{U}(1, 100)\) ms \\ Release \(\sim\boldsymbol{U}(50, 1000)\) ms}&  &  \checkmark & \checkmark \\ \hline
            Volume&  Gain \(\sim\boldsymbol{U}(-10, 10)\) dB&  &  \checkmark & \checkmark \\ \hline
            Low-Pass EQ& \makecell{Cutoff \(\sim\boldsymbol{U}(5512, 11025)\) Hz  \\ Roll-off = 6 db/octave} &  &  & \checkmark \\ \hline
            Hi-Pass EQ& \makecell{Cutoff \(\sim\boldsymbol{U}(32, 1024)\) Hz \\ Roll-off = 6 db/octave} &  &  & \checkmark \\ \hline
            Bitcrush& Bitrate \(\sim\boldsymbol{U}(2, 10)\)& & & \checkmark \\ \hline
            Delay& \makecell{Time \(\sim\boldsymbol{U}(10, 1000)\) ms \\ Feedback  \(\sim\boldsymbol{U}(0, 90)\)\% \\ Mix  \(\sim\boldsymbol{U}(10, 100)\)\%}& & & \checkmark \\ \hline
            Distortion& Drive\(\sim\boldsymbol{U}(1, 30)\) dB& & & \checkmark \\ \hline
            Convolution Reverb& \makecell{Impulse response \(\in\{1, ..., 14\}\) \\ Mix \(\sim\boldsymbol{U}(10, 100)\)\%} & & & \checkmark \\ \hline
            \multicolumn{2}{c|}{\textbf{Number of FX}} & 0 & $\in\{0, 1\}$ & $\in\{0, 1, 2, 3\}$ \\ 
        \bottomrule
        \end{tabular}
    \caption{Audio effects applied to ``sample'' stems when creating artificial ``mixtures''. Impulse responses for the ``reverb'' effect are from the \href{http://www.echothief.com/downloads}{EchoThief} library, with two IRs chosen at random from each of the seven provided categories.}
    \label{tab:stem_fx}
\end{table*}

\begin{table*}[t!]
    \centering
        \begin{tabular}{c|c|c|c|c}
        \toprule
            \textbf{FX name}&  \textbf{Parameters}&  \textbf{FX} \(\in\mathbb{A}\)&  \textbf{FX} \(\in\mathbb{B}\)& \textbf{FX} \(\in\mathbb{C}\)\\ \hline
            Pitch Shift& Transposition \(\in\{-3, -2, -1, 1, 2, 3\}\) semitones & \checkmark & \checkmark &  \\ \hline
            Microtone Shift& \makecell{Transposition \(\in\{-7, -6, -5, ..., 5, 6, 7\}\) semitones, \textbf{then:} \\ 10\% probability of transposing \(\sim\boldsymbol{U}(-0.5, 0.5)\) semitones} &  &  & \checkmark \\ \hline
            Time Stretch& Stretch factor \(\sim\boldsymbol{U}(70, 150)\)\% & & \checkmark & \checkmark \\ \hline
            Silence&  \multirow{4}{*}{\makecell{Frames-per-second \(\sim\boldsymbol{U}(0.5, 5.0)\) \\ Activation probability = 10\%}} & & \checkmark & \checkmark \\ \cline{1-1}\cline{3-5}
            Duplicate& & & \checkmark & \checkmark \\ \cline{1-1}\cline{3-5}
            Remove& & & \checkmark & \checkmark \\ \cline{1-1}\cline{3-5}
            Reverse& & & & \checkmark \\ \hline
            \multicolumn{2}{c|}{\textbf{Number of FX}} & $\in\{0, 1\}$ & $\in\{0, 1\}$ & $\in\{0, 1, 2, 3\}$ \\ 
        \bottomrule     
        \end{tabular}
    \caption{Audio effects applied to create artificial ``mixtures''.}
    \label{tab:mix_fx}
\end{table*}

To create the anchor clip, we apply a random selection of audio effects to the summed ``sample'' stems, including equalization, compression, and volume adjustment: the full range of effects is shown in Table \ref{tab:stem_fx}. The transformed ``sample'' is then summed together with the remaining ``non-sample'' stems and an additional layer of effects is applied. These effects include pitch and time stretching, as well as several designed to replicate a producer ``chopping'' a sample (Table \ref{tab:mix_fx}). Inspired by the time-warping augmentations in \citet{yesiler_accurate_2020}, for these effects we randomly choose a frames-per-second value between 0.5 and 5.0, which we use to chunk the audio mixture into non-overlapping frames. We then iterate over each frame sequentially and either duplicate, remove, silence, or reverse it (depending on the effect being applied) with 10\% probability. While it would be possible to add these effects solely to the sample stem rather than the entire mixture, to do so would likely result in a mismatched and unrealistic result with e.g., the sample stem transposed to a different key than the rest of the audio.

Alongside acting as data augmentation, these transformations also reflect common manipulations that music producers may apply to their samples \citep{chang_cant_2011}. To that end, we create three ``configurations'' of the artificial dataset, which we refer to as configurations $\mathbb{A}$, \(\mathbb{B}\), and \(\mathbb{C}\), and evaluate how each affects retrieval performance in our experiments. We vary the number and variety of effects between each configuration. Configuration $\mathbb{A}$ is the least extreme, with only a 50\% chance of applying pitch-shifting to the mixture of sample and non-sample audio, while $\mathbb{C}$ is the most extreme, with up to three different effects applied to both the sample and mixture.

The duration of the transformed positive clips is always 7.5 seconds; the duration of anchor clips varies, depending on the number of effects applied. To avoid the model simply learning to recognize the alignment between two clips (such as the position of a drum hit), we randomly crop the duration of both positive and anchor to 5 seconds. This ensures a minimum of 50\% overlap between both clips, while removing any exact alignment between them. In cases where the duration of a transformed anchor is under 5 seconds (only possible when the ``remove'' effect is used: see Table \ref{tab:mix_fx}), we use ``mirror'' (or reflect) padding to artificially extend the clip to the required duration.

We randomize the parameters used for each individual audio effect (as well as the number and types of effects applied) every epoch, which ensures that the model sees different transformations of the same input audio during training. Random numbers are sampled from uniform distributions, using the boundaries specified in Tables \ref{tab:stem_fx} and \ref{tab:mix_fx}. We make several examples of artificial positive and anchor audio clips available in our supplementary materials.\endnote{This will be released at a later date.}

\subsection{Proposed System}\label{sec:proposed_system}

\begin{figure*}[htbp]
  \centering
  \includegraphics[width=0.85\textwidth]{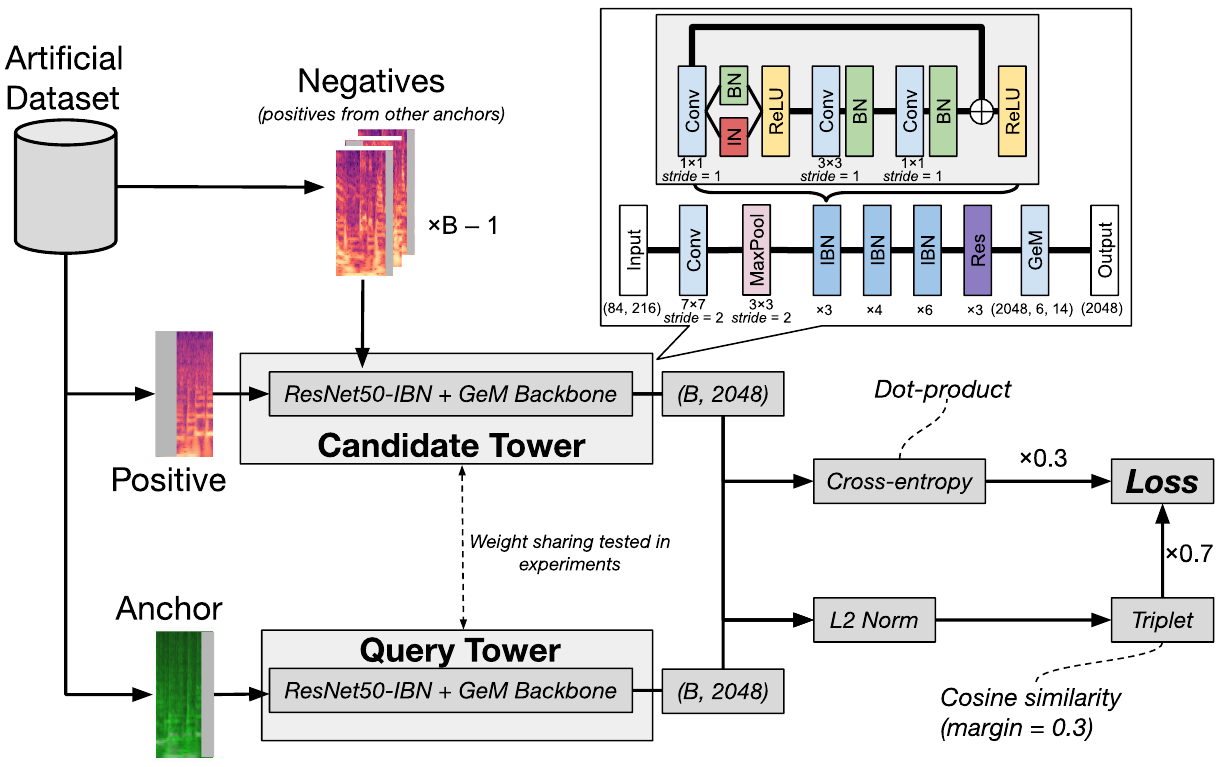}
  \caption{A diagram showing how the proposed system was trained. Anchor, positive, and negative audio clips in the artificial dataset are created following the procedure outlined in Figure \ref{fig:artificial_dataset}.}
\label{fig:training_diagram}
\end{figure*}

We show a diagram of the proposed system and training procedure in Figure \ref{fig:training_diagram}.

\subsubsection{Architecture}

We now outline the motivations behind the technical design of our system. 

We use a convolutional neural network for the encoder backbone of our architecture. CNNs possess a range of inductive biases that make them an effective choice for an initial attempt at designing a deep-learning sample identification system. In particular, the convolution operation is invariant to the translation of an input. This property allows the system to remain robust to common audio transformations such as moderate pitch-shifting, where the properties of the sample may shift, but the CNN can still recognize these patterns in the feature map. CNNs also have the capacity to detect local patterns through weight sharing and local receptive fields, which is useful to detect specific onsets, formants, and harmonic slopes, and which we hypothesize will help in recognizing key local characteristics of samples. 

More specifically, we use the ResNet50-IBN architecture as the encoder backbone. This is a modification of the classic ResNet50 architecture \citep{he_deep_2016}, where the first, second, and third residual modules are replaced with instance batch normalization (IBN) modules, leaving the final residual module unchanged. Prior computer vision research has demonstrated that these IBN layers assist in learning features that are robust to changes in the appearance and perspective of an image \citep{pan_two_2018}. Our motivation for using this architecture was based on the hypothesis that these IBN layers may also be able to learn representations of audio inputs that are robust to the changes in key, tempo, and timbre that are commonplace in sample-based music. Here, we note that this architecture has also been employed successfully by \citet{du_bytecover_2021} as part of a cover song identification system. In our experiments (see section \ref{sec:experiments}), we are able to demonstrate that the ResNet50-IBN outperforms a Transformer-based encoder.

We represent each audio clip in the artificial dataset using constant-Q transform (CQT) spectrograms with a hop size of 512 samples and 12 bins per octave. We found during our initial testing that using CQT spectrograms improved performance on the held-out artificial data over alternative input representations such as Mel spectrograms. 

After passing through the backbone, we compress the three-dimensional feature map from the final residual module using a generalized mean (GeM) pooling module, as in \citet{du_bytecover_2021}. This module has a single learnable parameter $p$, where the output is calculated as the $p$-th root of the average of the elements within the pooling region raised to the power of $p$. This results in a single one-dimensional embedding of size $d$ for every audio clip. We test the effects of different sizes of $d$ during our experiments (see section \ref{sec:experiments}) and find that $d=2048$ leads to optimal performance.

Rather than training a single ResNet50-IBN backbone as in \citet{du_bytecover_2021}, we instead employ a ``two towers'' structure, where weights are learned separately for two encoders --- one of which (the ``query tower'') embeds anchor audio clips and the other (the ``candidate tower'') embeds corresponding positive and negative clips. The motivation for this decision comes from several prior audio retrieval systems that use individual encoder towers to embed both anchor and positive examples with contrasting musical and instrumental qualities. For example, both \citet{zhang_siamese_2019} and \citet{zhang_learn_2021} found that training separate encoders to embed both isolated vocal recordings and original audio improved the generalization of the representation space compared to using a single encoder. In our experiments (see section \ref{sec:experiments}), we are able to demonstrate that training separate towers improves the performance of our system over training only a single tower.

\subsubsection{Multi-Loss Training}

\begin{figure*}[ht]
  \centering
  \includegraphics[width=0.85\textwidth]{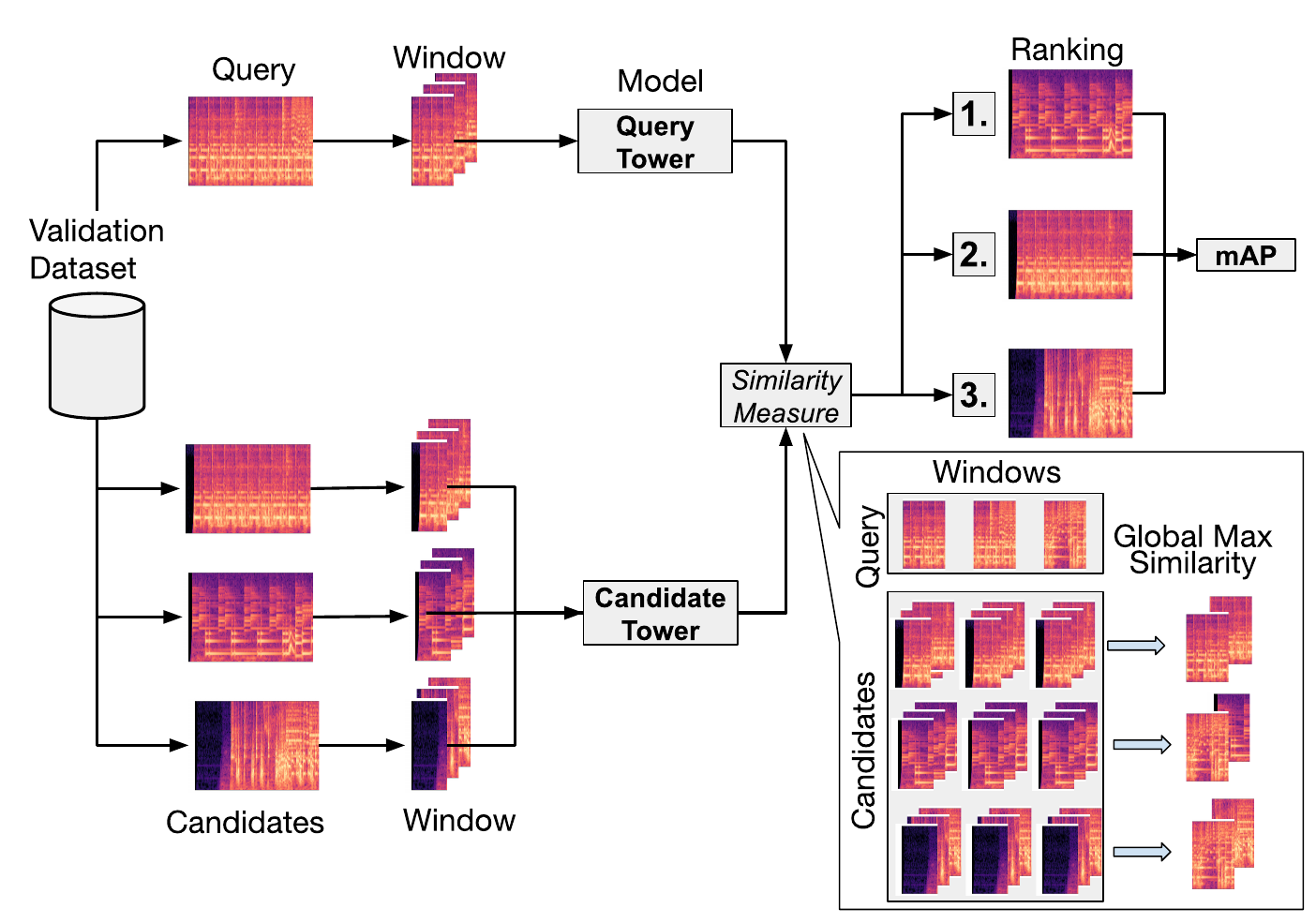}
  \caption{A diagram showing how the proposed system was used during evaluation. The query and candidate towers are trained using the artificial dataset as outlined in Figure \ref{fig:artificial_dataset}. During our experiments, we evaluate the effect of using only one tower to encode all audio, as well as the effect of changing the hop between extracted windows.}
\label{fig:inference_diagram}
\end{figure*}

We train the model separately to optimize both a classification and metric-learning loss function. The motivation behind using two separate losses is that the classification loss will be useful to help the model efficiently discriminate each known class (i.e., sample or non-sample), whereas the metric-learning loss will be useful to help the model generalize to unseen classes, by learning embeddings where similar content is naturally clustered together in the embedding space \citep{wang_jointly_2021}. We hypothesize that both losses will be important for sample identification, both to learn general rules differentiating between different types of samples (e.g., between drum breaks and vocal loops) and in learning subtle variations (e.g., different types of audio transformations). In our experiments (see section \ref{sec:experiments}), we are able to demonstrate that training with both losses improves performance over training with only a single loss.

The classification task involves discriminating between audio clips that either contain or do not contain the correct sample. For any given anchor audio clip, we treat the corresponding positive clip as the target class, and other positive samples from within the batch (but from different source music recordings) as negative classes, as in \citet{zhang_siamese_2019}.
We use the dot-product similarity of the non-normalized, one-dimensional vectors extracted from the final hidden layer of the model, and compute a softmax cross-entropy loss. For an anchor $C_a$ with corresponding positive $C_p$ and negatives $C^i_n$, $i = 1 \ldots N$,

\begin{align}\label{eq:cce_loss} 
L_{\text{cls}} = - C_p \cdot C_a + \log\sum_{c \in \{C_p\} \cup \{C^i_p\}} \exp(C \cdot C_a) \ .
\end{align}

The metric-learning task makes features belonging to clips with the same sample audio close together, and features from clips with different samples far apart in the embedding space. For a given anchor audio clip, we mine semi-hard negatives from within the batch (but, again, from different source recordings) within the margin $\alpha$. We take the $L_2$-normalization of extracted feature embeddings and then use a triplet loss to optimize their cosine similarity, such that
\begin{align}\label{eq:trip_loss}
L_{\text{trip}} = \text{max}\{0, \text{cos}(C_A, C_n) - \text{cos}(C_A, C_p) + \alpha\}, n = 1, ..., N \ . 
\end{align}

Accordingly, the final loss used to train the model can be written as:
\begin{align}\label{eq:final_loss}
L = \beta L_{\text{cls}} + \gamma L_{\text{trip}} \ ,
\end{align}
where $\beta$ and $\gamma$ are hyperparameters used to balance the contributions of the two losses, as is the case in \citet{wang_jointly_2021}. We test the effect of training only using either $L_{\text{cls}}$ or $L_{\text{trip}}$ during our experiments.

We train all models with the Adam optimizer for 100 epochs on NVIDIA A100 GPUs, with a batch size of 512 anchor-positive pairs. The initial learning rate is set to \(1e^{-4}\) and reduced by a factor of five at epochs 40 and 80. Following initial testing, we set the margin $\alpha$ used for mining triples from within a batch to 0.3, with the hyperparameters $\beta$ and $\gamma$ also set to 0.3 and 0.7, respectively. 

\section{Results}\label{sec:results}

\subsection{Evaluation Methodology}\label{sec:evaluation}

A diagram showing the evaluation procedure is given in Figure \ref{fig:inference_diagram}.

\subsubsection{Dataset}\label{sec:evaluation_dataset}

After training, we evaluate the performance of our model on the dataset from \citet{vanbalen_sample_2013}. This dataset consists of 76 full-length query music recordings that sample from 68 full-length candidate recordings a total of 137 times, with 104 unique query-candidate relations. In other words, some queries sample from multiple candidates, and some candidates appear in multiple queries. 

\begin{figure}[ht]
  \centering
  \includegraphics[width=0.85\columnwidth]{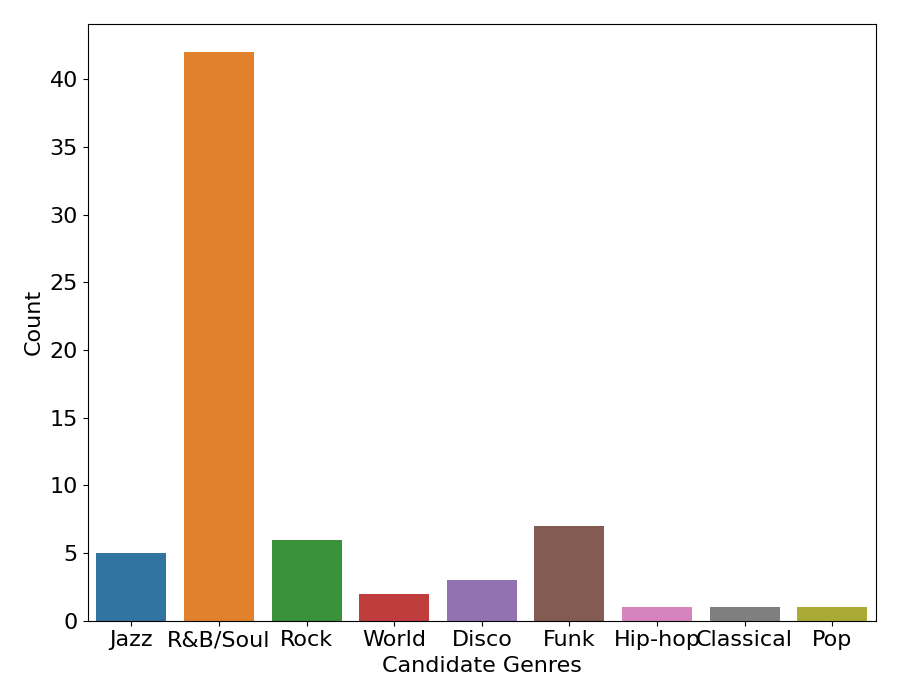}
  \caption{Distribution of candidate track genres in the commercial audio dataset. Each of the 68 candidate tracks are assigned a single genre.}
\label{fig:validation_genres}
\end{figure}

All queries are commercial hip-hop recordings while candidates come from a variety of genres, with the majority being R\&B/Soul recordings (Figure \ref{fig:validation_genres}). The query music recordings incorporate tonal and percussive (i.e., drum breaks) samples from the candidate pool, both with and without the use of pitch-shifting. Note that the candidate pool includes the full-length candidate music recordings, however only a small portion of each candidate is ever sampled in a query: the longest sample is twenty-six seconds long, while the shortest is one second. This is consistent with the desired use case of a sample identification system, where the exact portion of sampled audio may not be known, and is also consistent with the methodology established in previous work on this task \citep{gururani_automatic_2017, vanbalen_sample_2013, whitney_automatic_2013}.

In order to challenge the system, the candidate pool includes 320 additional ``noise'' tracks with a similar genre distribution to the actual candidates, but which are not sampled in any query. The initial description of this dataset did not explicitly identify the noise tracks that were added to the candidate pool, meaning that replicating the reported results was difficult. We make this information --- including both track titles and artist names --- available for the first time as part of our supplementary materials.\endnote{This will be released at a later date.}

\subsubsection{Similarity Measure}

In many cover song identification systems, query and candidate music tracks are each represented using a single feature vector, with the similarity between them calculated using a single operation (e.g., cosine, Euclidean distance). This is unhelpful in the sample identification task, because the portion of the query that is relevant to the candidates may only be a few seconds long. Consequently, we design a simple heuristic to compare a sequence of local feature vectors extracted from both a query and candidate music recording in the commercial audio dataset. 

For a given query \(\boldsymbol{X}\) and candidate \(\boldsymbol{Y}\), we extract sequences of $d$-dimensional local features from within a sliding window, such that \(\boldsymbol{X}\in\mathbb{R}^{M\times d}\) and \(\boldsymbol{Y}\in\mathbb{R}^{N\times d}\). Here, \(M\) and \(N\) represent the duration of \(\boldsymbol{X}\) and \(\boldsymbol{Y}\) and may be highly different, as in \citet{du_bytecover3_2023}. 

We then compute $\text{sim}(\boldsymbol{X}, \boldsymbol{Y})$ as the global maximum of pairwise cosine similarities between $\boldsymbol{X}$ and $\boldsymbol{Y}$:
\begin{align}\label{eq:global_max}
\text{sim}(\boldsymbol{X},\boldsymbol{Y}) &= \text{max}_i(\text{max}_j(\text{cos}(x_i, y_j)))
\end{align}
where $i = 1, ..., M$ and $j = 1, ..., N$.

By focusing only on the global maximum cosine similarity, we are able to ignore the contribution of features in a candidate that are irrelevant to a query. Our approach is conceptually similar to the work of \citet{du_bytecover3_2023}, who use the \textit{average} (as opposed to the maximum) of pairwise cosine maxima between short query recordings and full-length candidate cover versions. 

As with our artificial dataset, we set values for the size of the window used to extract local features to 5.0 seconds. We test the effect of changing the hop length between successive windows in our experiments.

\subsection{Experiments}\label{sec:experiments}

\begin{table*}
    \centering
        \begin{tabular}{c|c|c|c|c|c|c|c}
            \toprule
            \textbf{Tower} & \textbf{Backbone}     & \textbf{Dataset}   & \textbf{Stems} & \textbf{Hop Length}     & \textbf{dims.}      & \textbf{Loss}   & \textbf{mAP} $\uparrow$ \\ \hline
            \multicolumn{8}{c}{\textit{System Architecture}} \\ \hline
            Both towers         & \multirow{4}{*}{ResNet50-IBN} & \multirow{4}{*}{$\mathbb{B}$}& \multirow{4}{*}{$N-1$} & \multirow{4}{*}{2.5}& \multirow{4}{*}{2048}  & \multirow{4}{*}{Combined}& 0.334   \\ \cline{1-1} \cline{8-8}
            Shared weights      &                              &                               &                        &                      &                       & & 0.358   \\ \cline{1-1} \cline{8-8}
            Candidate tower only&                              &                               &                        &                      &                       & & 0.350   \\ \cline{1-1} \cline{8-8}
            Query tower only    &                              &                               &                        &                      &                       & & \textbf{0.441} \\ \hline
            \multicolumn{8}{c}{\textit{Encoder Backbone}} \\ \hline               
            Query tower         & ViT           & $\mathbb{B}$                  & $N-1$                  & 2.5                  & 2048                  & Combined & 0.267   \\ \hline       
            \multicolumn{8}{c}{\textit{Dataset Configuration \& Stems}} \\ \hline
            \multirow{3}{*}{Query tower} & \multirow{3}{*}{ResNet50-IBN} & $\mathbb{A}$         & \multirow{2}{*}{$N-1$} & \multirow{3}{*}{2.5} & \multirow{3}{*}{2048} & \multirow{3}{*}{Combined}& 0.398   \\ \cline{3-3} \cline{8-8}
                                          &                         & \multirow{2}{*}{$\mathbb{C}$} &                    &                      &                       & & 0.394   \\ \cline{4-4} \cline{8-8}
                                          &                         &                               & $N$                &                      &                       & & 0.363   \\ \hline
            \multicolumn{8}{c}{\textit{Hop Length}} \\ \hline
            \multirow{3}{*}{Query tower} & \multirow{3}{*}{ResNet50-IBN} & \multirow{3}{*}{$\mathbb{B}$} & \multirow{3}{*}{$N-1$} & 1.0         & \multirow{3}{*}{2048} & \multirow{3}{*}{Combined} & 0.408   \\ \cline{5-5} \cline{8-8}
                                          &                         &                               &                      & 4.0                &                      &  & 0.391   \\ \cline{5-5} \cline{8-8}
                                          &                         &                               &                      & 5.0                &                      &  & 0.383   \\ \hline
            \multicolumn{8}{c}{\textit{Embedding Dimension}} \\ \hline
            \multirow{2}{*}{Query tower} & \multirow{2}{*}{ResNet50-IBN} & \multirow{2}{*}{$\mathbb{C}$} & \multirow{2}{*}{$N-1$} & \multirow{2}{*}{2.5} & 512          & \multirow{2}{*}{Combined}& 0.359   \\ \cline{6-6} \cline{8-8}
                                          &                         &                               &                      &                      & 8192               &  & 0.361   \\ \hline
            \multicolumn{8}{c}{\textit{Loss Function}} \\ \hline
            \multirow{2}{*}{Query tower} & \multirow{2}{*}{ResNet50-IBN} & \multirow{2}{*}{$\mathbb{B}$} & \multirow{2}{*}{$N-1$} & \multirow{2}{*}{2.5} & \multirow{2}{*}{2048}          & Cross-Entropy & 0.380 \\ \cline{7-8}
            & & & & & & Triplet & 0.209 \\
            \bottomrule
        \end{tabular}
    \caption{Experiment results. Dataset refers to the number and type of effects applied to the audio, with $\mathbb{A}$ being the least and $\mathbb{C}$ the most extreme. Stems refers to the number of audio files that are treated as the ``sample'' out of $N$ extracted using source separation, where $N\in \{2, 3\}$ and $1 < N < 4$. Hop length is the time between successive local feature windows (in seconds), and dims is the size of the extracted features.}
    \label{tab:experiment_results}
\end{table*}

The results of each experiment are described below and presented in Table \ref{tab:experiment_results}. To report performance, we follow \citet{vanbalen_sample_2013} in using the mean average precision (mAP) of the results returned by each system or model; for every query, we rank each item in the candidate pool using our similarity measure, calculate the precision of each relevant candidate sample, average the results, and finally take the mean of all average precision scores. Unless stated otherwise, all audio in the commercial dataset is downmixed to mono with a sample rate of 22,050 Hz (as described in section \ref{sec:audio_sources}).

\subsubsection{Effect of System Architecture}

We begin by comparing our ``two towers'' architecture to an architecture where weights are shared between both towers. In both cases, the ``query tower'' (trained on artificial mixtures) is used to embed query recordings from the dataset, and the ``candidate tower'' (trained on artificial samples) embeds candidate samples. We find that sharing weights between these towers marginally increases performance over training both separately, with a difference in mAP of approximately $+0.02$. This might suggest that the model benefits from learning a unified representation space for both samples and mixtures, over learning features separately for each.

Interestingly, however, we find that the performance of the model improves much more substantially when training both towers separately, but only using \textit{one} of these to embed audio from the commercial dataset. This increase is particularly significant when only the query tower is used (with the candidate tower discarded after training), with an improvement of 32\% over using both towers during evaluation ($\text{mAP} = 0.44$, compared with $0.33$). One explanation is that the real-world candidate recordings have more in common with the mixture clips in our artificial dataset than the sample clips; they typically contain \textit{vocals}, \textit{harmony}, and \textit{drums} elements together, rather than only a subset of these. Another possibility is that the candidate tower functioned as a surrogate to help the query tower learn a flexible representation space. 

\subsubsection{Effect of Encoder Backbone}

We test the effect of replacing the ResNet50-IBN backbone used in both towers with a Vision Transformer (ViT) model. The ViT has 12 transformer layers and 12 attention heads, following the work of \citet{gong_ast_2021} in using these models for audio fingerprinting tasks. We use a a CQT spectrogram as the input to the ViT with dimensions (224, 224), yielding (16, 16) ``patches'' as in \citet{dosovitskiy_image_2021}. For parity with the ResNet50-IBN backbone, we compress the output of the final hidden state to a one-dimensional feature vector with $d = 2048$ using a fully-connected layer and GeM pooling. 

The weights of the ViT are initialized from pre-training on ImageNet-21k \citep{deng_imagenet_2009}; this was demonstrated to outperform random initialization of weights in prior audio fingerprinting work \citep{gong_ast_2021}. We then remove the classification head and train the model for 50 epochs using dataset configuration $\mathbb{B}$, with a batch size of 128.\endnote{Note that both the number of training epochs and the batch size are lower for the ViT here compared with our ResNet50-IBN, in order to account both for the pre-training and the increased size of the transformer model. However, from observing the artificial validation loss, we believe the ViT successfully converged within this time, and we do not think it would perform significantly better with more training.} All other training hyperparameters are maintained between both encoders, and we train both towers but use only the query tower during evaluation. 

The ViT significantly underperformed compared to the ResNet50-IBN system trained with the same combination of parameters, only achieving $\text{mAP} = 0.27$ --- approximately 40\% worse than the best-performing ResNet50-IBN. One possible explanation is that the amount of artificial training data may not have been sufficient to optimize performance of the ViT; previous research has found that transformers typically only outperform convolutional architectures at larger scales than we tested here \citep{dosovitskiy_image_2021}.

\subsubsection{Effect of Dataset Configuration}

We evaluate the three combinations of audio effects used to create artificial training examples, referred to as configurations $\mathbb{A}$, $\mathbb{B}$, and $\mathbb{C}$ in Table \ref{tab:stem_fx} and \ref{tab:mix_fx}. We also test a variant of dataset configuration $\mathbb{C}$ where up to $N$ stems from a recording are treated as the ``sample'' (as opposed to $N - 1$: see section \ref{sec:window_extraction}), meaning that a thirty-second recording yields 70 anchor-positive pairs. 

We find that dataset configuration $\mathbb{B}$ performs the best in identifying real-world instances of sampling. Somewhat surprising is that configurations $\mathbb{A}$ and $\mathbb{C}$ obtain similar precision ($\text{mAP} = 0.40$ and $0.39$, respectively), despite substantial differences in the number and variety of effects applied. One possibility is that configuration $\mathbb{B}$ achieved a ``Goldilocks'' effect, applying just enough transformation to the training audio to force the model to learn robust feature representations without over-fitting to these transformations. Alternatively, configuration $\mathbb{C}$ may be too extreme for the types of transformation applied to the commercial hip-hop samples; it is possible that it may perform better for forms of electronic and dance music sampling, where extensive transformation and ``chopping'' of a sample are commonplace \citep{schloss_making_2013}.

Finally, the variation of dataset configuration $\mathbb{C}$ using $N$ stems performed worse than the equivalent configuration using $N - 1$ stems ($\text{mAP} = 0.36$ and $0.39$, respectively). One simple explanation is that transforming all of the stems in a mixture may encourage the model to overfit to these transformations. Another explanation is that hip-hop producers may be more likely to add their own material to a sample rather than using it outright wholesale.

\subsubsection{Effect of Hop Length}

We also test the effect of changing the hop length $h$ between successive windows used to extract local features from the commercial audio dataset, setting $h \in \{1.0, 2.5, 4.0, 5.0\}$ seconds (equivalent to 80, 50, 20, and 0\% overlap between successive windows). Note that the hop length used for the artificial training data is not changed, with all models using a hop length varying between 2.5 and 5.0 seconds due to the random cropping of audio clips (see section \ref{sec:audio_fx}). We find that setting $h = 2.5$ seconds results in the best performance ($\text{mAP} = 0.44$), with values above this leading to reductions in precision; where $h = 5.0$, $\text{mAP} = 0.38$.

\subsubsection{Effect of Embedding Dimension}

In prior cover song identification work \citep[e.g.,][]{yesiler_accurate_2020, chang_neural_2021, du_bytecover2_2022, yesiler_less_2020}, increasing the size of the embedding dimension has typically improved the performance of a retrieval system, albeit with diminishing returns above a certain threshold and increases in processing time. Consequently, in this experiment, we study the effect of the size of the embedding dimension $d$, training separate models on dataset $\mathbb{C}$ (with $N - 1$ stems) and setting $d \in \{512, 2048, 8192\}$.

We observe that the performance of our proposed approach is strongest with $d = 2048$ ($\text{mAP} = 0.44$), with both the larger and smaller embedding sizes performing substantially worse (both $\text{mAP} = 0.36$). The smaller embedding may not have contained enough information to effectively distinguish between samples, while the larger embedding may have introduced redundancy --- with not enough information contained in the artificial audio clips for meaningful feature representations to be learned. An alternative possibility is that the amount of training data may not have been sufficient to optimize the model with the largest embedding size.

\subsubsection{Effect of Loss Function}

Finally, we test the effect of training the model with only a single (either cross-entropy or triplet) loss. When training with the cross-entropy loss $L_{\text{cls}}$, we set the constant terms (previously used to balance the two losses) $\beta = 1$ and $\gamma = 0$; vice-versa, when training with the triplet loss $L_{\text{trip}}$ only, we set $\beta = 0$ and $\gamma = 1$. In both cases, two ResNet50-IBN towers are trained on dataset $\mathbb{B}$ with $d = 2048$ and $h = 2.5$ and the query tower used during inference. 

Training solely with the cross-entropy loss led to approximately 58\% better performance than training solely with the triplet loss ($\text{mAP} = 0.380$ vs. $0.209$). This might suggest that it is more helpful for the model to learn to classify between sample and non-sample audio than map the former closer in the embedding space to a query. However, training with both losses combined outperformed both single loss models, performing approximately 71\% better than the triplet loss only model and 15\% better than the cross-entropy loss only. This validates our decision to train with a combined loss, and also underscores how it may be helpful for the model to learn to optimize for \textit{both} inter-class discrimination and intra-class compactness \citet{du_bytecover_2021}.

\subsubsection{Overall Effects}

We obtain the best results ($\text{mAP} = 0.44$) for our proposed system by training two ResNet50-IBN towers on artificial dataset configuration $\mathbb{B}$ using a combined cross-entropy and triplet loss, with $N - 1$ stems and embedding dimension $d = 2048$. During evaluation, we use a hop length of $h = 2.5$ seconds between local features and embed all audio from the commercial dataset with the query tower (discarding the candidate tower). 

With these parameters, our proposed system retrieves a correct sample at rank 1 for a a total of 33 out of 76 queries (43.4\%), increasing to 40 songs by rank 5 (52.6\%). Both percussive and tonal samples are correctly retrieved by our model, alongside several pitch-shifted samples. Remarkably, in several instances the model correctly retrieves samples pitch-shifted outside the range of semitone values seen during training; for example, the sample of ``Cold as Ice'' by Foreigner (\citeyear{foreigner_cold_1977}) used in the song of the same name by M.O.P. (\citeyear{mop_cold_2000}) is transposed up five semitones, but still retrieved correctly by our model. 

\subsubsection{Visualization of Embedding Space}

\begin{figure}[ht]
  \centering
  \includegraphics[width=0.85\columnwidth]{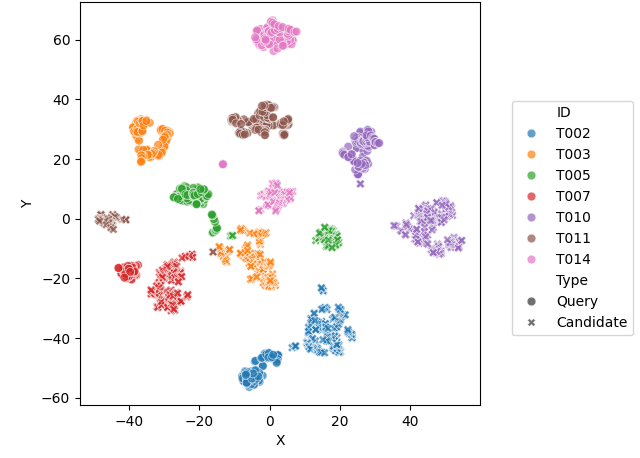}
  \caption{t-SNE plot of embedding features from a subset of commercial recordings. Recordings that contain the same sample are shown in the same color, while queries and candidates are shown using dot and cross markers, respectively. Track IDs refer to those given in \citet{vanbalen_sample_2013}.}
\label{fig:tsne_fig}
\end{figure}

To qualitatively evaluate the representations learned by our best-performing model, in Figure \ref{fig:tsne_fig} we visualize the embeddings of fourteen commercial tracks (seven queries sampling from seven candidates) from the evaluation dataset using the t-SNE method. In most cases, embeddings from the same audio recording are nicely clustered together. We can also begin to see overlap between embeddings from several queries and candidates that contain the same underlying sample (e.g., \texttt{T002}, \texttt{T007}).

However, this figure also illustrates inherent difficulties in the sample identification task as, in a number of cases, the apparent overlap between query and candidate is clearly very small. For instance, in both \texttt{T005} and \texttt{T010} only a single candidate window is mapped closely to the cluster of query embeddings, potentially indicating a situation where only a brief portion of the candidate is looped throughout the query. Vice-versa, for \texttt{T014} only a single query window is mapped near a cluster of candidate embeddings: this could indicate a situation where the candidate sample appears only once in the query (e.g., as a ``one shot'' sample).

\subsection{Baseline Methods}

\begin{table}[htbp]
    \centering
        \begin{tabular}{c|c}
        \toprule
            \textbf{System} & \textbf{mAP} $\uparrow$ \\ \hline
            \multicolumn{2}{c}{\textit{Acoustic Landmarks} \citep{wang_industrial-strength_2004}} \\ \hline
            Default settings \citep{ellis_audfprint_2014} & 0.050 \\ \cline{2-2}
            + optimizations \citep{vanbalen_sample_2013}  & 0.218 \\ \cline{2-2}
            + re-pitching \citep{vanbalen_sample_2013} & 0.390 \\ \hline
            \multicolumn{2}{c}{\textit{Panako} \citep{six_panako_2014}} \\ \hline
            Our modifications to \citet{six_panako_2021} & 0.009 \\ \hline
            \multicolumn{2}{c}{\textit{This Work}} \\ \hline
            Best result from Table \ref{tab:experiment_results} & \textbf{0.441} \\
        \bottomrule
        \end{tabular}
    \caption{Comparison of the best performing model in Table \ref{tab:experiment_results} against the baseline systems.}
    \label{tab:baseline_results}
\end{table}

We compare our best performing system with two audio fingerprinting systems described by  \citet{wang_industrial-strength_2004} (referred to here as \textit{Acoustic Landmarks}) and \citet{six_panako_2021} (referred to as \textit{Panako}), with the results shown in Table \ref{tab:baseline_results}. For more recent (non-fingerprinting) systems used in automatic sample identification \citep[e.g.,][]{whitney_automatic_2013, gururani_automatic_2017}, either the code or original evaluation data was not available. 

The \textit{Acoustic Landmarks} fingerprinting algorithm works by extracting pairs of promising spectral peaks (``landmarks'') from candidate recordings and storing these in a hash table. Given a query track, identical landmarks are retrieved from the table and matches are found in cases where the time differences between a substantial number of successive landmarks are the same \citep{wang_industrial-strength_2004}. The first variant of this system we test simply uses the default settings found in the implementation of the algorithm by \citet{ellis_audfprint_2014}.

\citet{vanbalen_sample_2013} introduced several optimizations of the \textit{Acoustic Landmarks} system for the task of sample identification. Briefly, they modified the density, spacing, and number of spectral peaks computed per landmark, changed parameters of the input audio (including downsampling it to 8,000 Hz), and re-pitched landmarks extracted from query recordings by up to $\pm$ 2.5 semitones (in 0.5 semitone steps) to account for transposition of the sample. Their optimizations were performed to maximize retrieval precision on the same dataset of commercial music recordings used in this work; as such, we report their obtained results (both with and without the re-pitching of query landmarks) to reflect the potential of the \textit{Acoustic Landmarks} system for this task.

The \textit{Panako} fingerprinting algorithm has similarities with the \textit{Acoustic Landmarks} system; local maxima are identified from a spectral representation of an audio input obtained using a constant-Q transform, stored in a hash table, and matched with equivalent fingerprints from a query track. However, unlike this system, \textit{Panako} is explicitly designed to be indifferent to transformations of a candidate within a query. This is accomplished by computing fingerprints as triplets of time-frequency events; thus, invariance to time-shifting can be achieved by storing only the \textit{ratios} between the time differences of events within the triplet (as opposed to the absolute difference), and invariance to pitch-shifting by storing only the frequency differences \citep{six_panako_2014}.

We make a small number of modifications to the implementation of \textit{Panako} described in \citet{six_panako_2021} for the purpose of sample identification. The default configuration only registers a match when 20\% of a candidate matches with a query, which we reduce to 1\% (noting that a five-second sample in a song lasting 210 seconds corresponds to approximately a 2\% match). We also reduce the number of matching fingerprints required to register a match from 10 to 1, again to account for cases where only a small part of a candidate appears in a query. All other parameters are set to the defaults specified by the authors. Our modified version of the \textit{Panako} system can be found on a separate repository.\endnote{\url{https://github.com/HuwCheston/Panako-SampleID}}

Our proposed deep-learning approach outperforms both of the baseline fingerprinting systems. In particular, \textbf{our system achieves 13\% greater precision than the best results previously obtained for this dataset} by \citet{vanbalen_sample_2013} (given in Table \ref{tab:baseline_results} as optimizations + re-pitching). This demonstrates that deep neural networks have the capacity to outperform traditional audio signal processing in sample identification tasks. A final advantage of our approach over this system is that it works using the ``raw'' query audio as input, and does not require this to be manually re-pitched prior to computing the best match. 

Given the claimed invariance towards pitch- and time-stretching, the performance of the \textit{Panako} system was disappointing --- performing significantly worse than the \textit{Acoustic Landmark} system even prior to optimization. It is possible that the system is not robust towards cases where pitch and time stretching are present alongside additional spectral noise, such as that introduced when new material or other samples are overlaid against an initial candidate to create a new composition.

\subsection{Sample Location}\label{sec:sample_location}

\begin{table}[htbp]
    \centering
    \begin{tabular}{c|c|c|c|c}
        \toprule
            $k$s & $\pm 2.5$&  $\pm 5.0$& $\pm 7.5$ & $\pm 10.0$ \\ \hline
            \textbf{mAP} $\uparrow$ & 0.413 & 0.476 & 0.519 & 0.550 \\
        \bottomrule
    \end{tabular}
    \caption{Results for the sample location task, obtained using the best performing model in Table \ref{tab:experiment_results}}
    \label{tab:location_results}
\end{table}

As a final test of the generality of the learned embeddings, we evaluate how well our best-performing system is able to retrieve the location of a sample, given both a candidate and a query that contains it. 

Our commercial audio dataset contains 137 individual instances of sampling between a query and candidate, which includes cases where a query samples several times from the same candidate. For each of these sampling instances, we extract local features from the candidate and calculate the maximum cosine similarity $s$ with equivalent features extracted from the query. This meant that, for a candidate with duration $N$, we obtain $\{s_i\}^N_{i=1}$. We sort these values from most to least similar and assign timestamps according to the beginning of the feature extraction window, setting the hop length $h = 2.5$s as before.

We compare these values with the corresponding annotated timestamps provided by \citet{vanbalen_sample_2013}, and use the mAP @ $\pm k$ seconds as the evaluation metric. Thus, if the timestamp identified by the model is within $\pm k$ seconds of the annotation, it is counted as a hit, while all other values are counted as misses.\endnote{It is worth noting that if a sample appears several times \textit{exactly} in a candidate (such as through the use of looping), this could increase the false negative rate as the annotated timestamps would only refer to one of these instances. However, given that the genre distribution of tracks in the commercial music dataset favors of music that would typically be recorded ``live'' (e.g., R\&B/Soul and Jazz; see Figure \ref{fig:validation_genres}) and where post-hoc manipulation of a recording is rare, it is unlikely that the sampled material would appear exactly the same in these cases.} We obtain the average precision for a given query-candidate relation as the average of all times the candidate appears in the query, and compute the mean of these results across every query-candidate relation in the dataset as the mAP. We use values of $k \in \pm \{2.5,5.0,7.5,10.0\}$ seconds to test the sensitivity of our system (Table \ref{tab:location_results}). 

In total, our model correctly locates the position of the sample within $\pm 5.0$ seconds for approximately half of the sampling instances in the commercial music dataset. Precision improves somewhat as values of $k$ increase, with an mAP @ $k = \pm10.0$ seconds of $0.55$.

\begin{figure*}[htbp]
  \centering
  \includegraphics[width=0.85\textwidth]{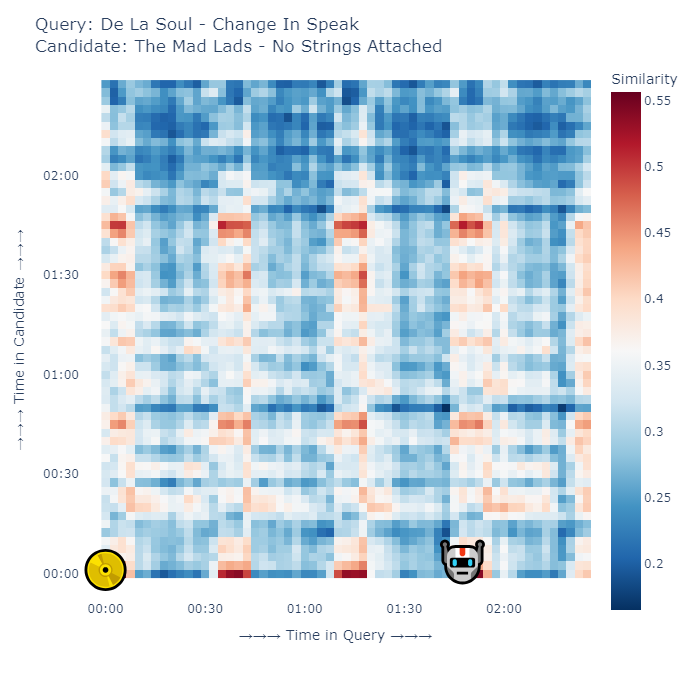}
  \caption{Pairwise cosine similarity values obtained for a single query and candidate recording in the dataset of commercial samples, using the best performing model from Table \ref{tab:experiment_results}. Time in the query \citep{delasoul_changeinspeak_1989} is shown along the $x$-axis, and along the $y$-axis for the candidate \citep{themadlads_nostringsattached_1969}, with darker red colors indicating greater similarity between the two recordings. The ground truth timestamp from \citet{vanbalen_sample_2013} is shown using the record symbol, with the largest cosine similarity value estimated by the model shown with the robot symbol.}
\label{fig:sample_heatmap}
\end{figure*}

In Figure \ref{fig:sample_heatmap}, we show a heatmap of the cosine similarity between the pairwise combinations of local features extracted from ``Change in Speak'' \citep{delasoul_changeinspeak_1989}, which contains a sample of ``No Strings Attached'' \citep{themadlads_nostringsattached_1969}. The predicted location of the sample is indicated by a robot symbol on the plot. We make similar plots available for every instance of sampling contained in the commercial audio dataset as part of an interactive web application included in our supplementary materials.\endnote{This will be released at a later date.}

\section{Discussion}

This paper introduces what what we believe to be the first deep learning model capable of the task of retrieving audio samples from hip-hop music. We train two convolutional neural networks on an artificial dataset of ``fake'' samples that are extracted from non-commercial music recordings and transformed using digital effects. Through several experiments, we demonstrate that (despite never having seen an ``actual'' samples during training) our model outperforms several baseline audio fingerprinting systems designed to be robust towards audio transformation. We also demonstrate that our system is capable of generalizing to the task of identifying where a sample appears in a given recording. More broadly, our results show how machine listening models can perform audio retrieval tasks previously reserved for experts.

There are several possible applications of this work, including as a tool to assist musicologists or other researchers in better quantifying or understanding music containing samples \citep{vanbalen_sample_2013}. Another possible application is in domain-specific, content-based music replacement tasks. Typically, samples used in commercial hip-hop recordings are ``cleared'' with the original copyright holders, and unlicensed sampling has led to several high-profile legal cases \citep{dittmar_audio_2012, chang_cant_2011}. Our system could be used to replace a sample that a producer cannot obtain clearance for with a similar one that they are able to license. This would require computing the distance between embeddings extracted from a query track containing an uncleared query sample and a pool of other (cleared) candidate samples, with the minimum distance treated as a potential replacement (see, for instance, Figure \ref{fig:tsne_fig}). We note here that metric learning approaches similar to those adopted in this work have often been used in music replacement scenarios \citep{lee_disentangled_2020, lee_similarity-based_2021}.

We can think of several limitations of our current work. Primarily, while a mAP of 0.441 represents a significant improvement on the baseline fingerprinting systems --- and is comparable with early deep learning systems for cover song identification \citep{xu_key-invariant_2018} --- it is still rather low for a retrieval task. While this could be due to the inherent difficulty of the sample identification task, it is also worth speculating why this might be the case for our system in particular. 

Firstly, from listening to several of the samples in the commercial audio dataset that the model failed to retrieve, it became apparent that they may not have been directly sampled in a query but ``interpolated'' --- i.e., recreated from scratch by the music producer --- instead.\endnote{For an example, see the sample of ``(You) Got What I Need'' \citep{scott_yougotwhatineed_1968} in ``Just a Friend'' \citep{markie_justafriend_1989}. Note, however, that whether material is sampled or interpolated is difficult to confirm conclusively.} The artificial anchor and positive examples we use during training came from the same underlying source recording, making the task conceptually similar to audio fingerprinting. It is possible that instead using two \textit{versions} of an underlying work (possibly with transformations) --- closer, in other words, to the cover song identification task --- may improve detection of these samples.

A second, related, limitation of the current work is the possibility that the model could have learned contextual ``shortcuts'' for mapping artificial candidates and queries, that were not generalizable to commercial hip-hop samples. These could include, for instance, artifacts that bled through the source separation model, such as the remnants of a drum hit appearing in the vocal stem, when the latter was treated as the positive example and the former as part of the anchor. An alternative approach to creating an artificial dataset might involve combining stems (or other audio material, such as sound effects) from \textit{different} sources together to create mixtures. However, careful attention would need to be given to making sure that these sources are directly compatible with each other in order for the artificial examples to represent real-world sampling \citep{huang_modeling_2021}, which we sidestep by using material from the same source.

A third limitation is that it is hard to be certain whether the transformations we apply to our artificial samples during training are realistic. For instance, while it is evident through listening that effects like equalization and pitch shifting are often applied to hip-hop samples, knowing exactly which parameter settings are typically used (e.g., filter position and depth, amount of shifting) is a different story. Previous work has successfully reconstructed the effects applied to electric guitar recordings \citep{jurgens_recognizing_2020}, as well as the parameter settings used on analog synthesizers \citep{yee-king_automatic_2018}. If a similar system could be adapted to identify the effects applied to a sample, a more targeted and realistic suite of manipulations than the ones employed here could be designed. 

A final limitation of this work is that the model architecture we employed may not be as suitable for sample identification as it is for cover song identification. It is interesting that the ResNet50-IBN backbone performs substantially better at cover song identification than sample identification; for the ``Covers80'' dataset \citep{ellis_covers80_2007}, equivalent in size to our dataset of commercial audio, \citet{du_bytecover_2021} achieved an mAP approximately twice that reported here. What this could highlight is that cover song identification may not be directly comparable with sample identification. Domain specific system architectures may be worth exploring, instead; for instance, the spectrogram patches used as tokens to a ViT model could be centered around prominent spectral landmarks, which may help it to focus solely on the elements of a sample that are most relevant to a new song.

Future work in this area would benefit from a larger dataset of real-world instances of sampling (equivalent to those typically used in audio fingerprinting and cover song identification), which could be used in training and evaluating models at a larger scale to that explored here. Our commercial audio dataset consists of 137 instances of sampling, meaning that it is comparable with smaller cover song identification datasets like \citet{ellis_covers80_2007} --- but vastly smaller than others like \citet{yesiler_da-tacos_2019}, which includes over 10,000 recordings. One complication to creating a larger dataset is that, while online services that link samples to songs that incorporate them do exist, this information is provided by a community of users, which makes verification difficult. A human-in-the-loop system could be designed to assist with this, using a model such as the one described here to obtain a baseline indication of the similarity between an identified query and candidate music recording, with ambiguous results flagged for manual checking.

Taken together, this paper describes what we believe to be the first neural sample identification model designed for hip-hop music. Much work remains to improve the accuracy and precision of our results; however, we hope that the public release of the artificial dataset and training code used in this work will help aid this process.

\section{Reproducibility}


The code used to both create the artificial dataset and train the model will be made available at a later date. This also extends to our electronic supplementary materials, including documentation, audio examples, and an interactive web application for exploring similar graphs to those contained in Figure \ref{fig:sample_heatmap} for
every recording. We make the modified version of the Panako audio fingerprinting system available at \url{https://github.com/HuwCheston/Panako-SampleID}


\section{Competing interests}

The authors declare that they have no competing interests.




\IfFileExists{\jobname.ent}{
   \theendnotes
}{
}

\section*{Acknowledgments}


The authors wish to express their thanks to Peter Sobot for technical assistance with this project and to Sebastian Ewert for helpful comments on an initial draft of this paper.



\printbibliography

\end{document}